# 300 K Ferromagnetic Magnetic Circular Dichroism in $Co^{2+}$-doped ZnO Activated by Shallow Donors

Kevin R. Kittilstved, Jialong Zhao, William K. Liu, J. Daniel Bryan, and Daniel R. Gamelin*

*Department of Chemistry, University of Washington, Seattle, WA 98195-1700*

Cobalt-doped ZnO ($Co^{2+}$:ZnO) films were studied by magnetic circular dichroism (MCD) spectroscopy. A broad 300 K ferromagnetic MCD signal was observed between 1.4 and 4.0 eV after introducing shallow donor states by exposure of paramagnetic $Co^{2+}$:ZnO films to zinc vapor. The new sub-bandgap ferromagnetic MCD intensity is attributed to low-energy photoionization transitions originating from a spin-split donor impurity band in ferromagnetic $Co^{2+}$:ZnO.

Diluted magnetic semiconductors (DMSs)[1] have attracted much recent attention due to their potential application in semiconductor spintronic devices.[2] For practical applications, ferromagnetic DMSs with Curie temperatures ($T_C$) well above room temperature are desired. Important advances have been made by several laboratories toward understanding the high-$T_C$ ferromagnetism observed in oxide DMSs.[3,4] In some cases, dopant clustering was observed and the ferromagnetism evidently had extrinsic origins.[5] In other cases, mounting evidence points toward intrinsic ferromagnetism that relies on the presence of $n$- or $p$-type lattice defects.[3,4,6] $Co^{2+}$:ZnO grown without deliberate incorporation of such defects has been studied extensively and exhibits only paramagnetism (with weak local antiferromagnetic superexchange),[7,8] but weak ferromagnetism can be induced by vacuum annealing,[8] and stronger ferromagnetism can be achieved by growth under lower $O_2$ partial pressures,[4,9] both linked to introduction of shallow donor defects. Quantitatively reversible 300 K ferromagnetic ordering in $Co^{2+}$:ZnO was demonstrated by Zn vapor diffusion ("on") and subsequent aerobic re-oxidation ("off"), attributed to introduction and removal of the shallow donor, interstitial zinc ($Zn_i$).[10]

MCD spectroscopy provides a sensitive tool for exploration of the magnetic and electronic structural properties of DMSs. MCD measurements at the band edge of paramagnetic $Co^{2+}$:ZnO have verified the existence of a giant excitonic Zeeman effect,[11,12] analysis of which yielded a $p$-$d$ exchange energy of $N_0\beta \approx -2.3$ eV.[12] To date, however, the MCD spectroscopy of ferromagnetic oxide DMSs has received relatively little attention.[13-15] In these studies, very broad MCD features were observed that generally bore little resemblance to those of the parent paramagnetic DMSs. This discrepancy was invoked to suggest extrinsic ferromagnetism in $Co^{2+}$:ZnO.[13] In $Co^{2+}$:$TiO_2$, however, similar broad MCD spectra were interpreted as arising from intrinsic DMS ferromagnetism,[14,15] but the electronic structural origins of the sub-bandgap MCD intensities were not addressed. In this Letter we describe the observation of similar broad sub-bandgap MCD intensities in $Co^{2+}$:ZnO films exposed to Zn vapor. This intensity is attributed to photoionization transitions originating from a spin-split donor impurity band, supporting proposals of intrinsic donor-mediated ferromagnetism in $Co^{2+}$:ZnO.

Figure 1a shows 300 K absorption spectra of a ~50 nm thick 4.2% $Co^{2+}$:ZnO film before (dashed) and after (solid) exposure to Zn vapor at 450 °C for 1 hr.[16] The absorption spectrum of the as-prepared film is the same as those of paramagnetic $Co^{2+}$:ZnO single crystals, showing the characteristic spin-orbit split $^4A_2 \rightarrow {}^4T_1(P)$ ligand-field transition of substitutional $Co^{2+}$ ions at ~2.0 eV (see inset)[17] and the first excitonic absorption maximum of ZnO at ~3.5 eV. Upon exposure to Zn vapor, the ZnO absorption edge shifted slightly to higher energy and a broad, weak absorption tail was observed throughout the visible energy range. Under some conditions, a new absorption band could be detected at ~0.4 eV, as described previously.[10] These spectroscopic changes were also observed in undoped ZnO exposed to Zn vapor and are characteristic of the introduction of shallow donors, in this case likely $Zn_i$.[10,18,19]

Figure 1b shows 300 K, 0.5 T MCD spectra of the same $Co^{2+}$:ZnO film before and after exposure to Zn vapor. Before Zn exposure, only very weak MCD intensity was observed at 300 K (dashed line in Fig. 1b). Magnification of this MCD spectrum reveals the $Co^{2+}$ $^4A_2 \rightarrow {}^4T_1(P)$ band, which agrees well with that observed by absorption (Fig. 1a inset). Variable-temperature, variable-field MCD spectroscopy and SQUID magnetometry confirmed that these $Co^{2+}$:ZnO films were paramagnetic, with strongly temperature-dependent

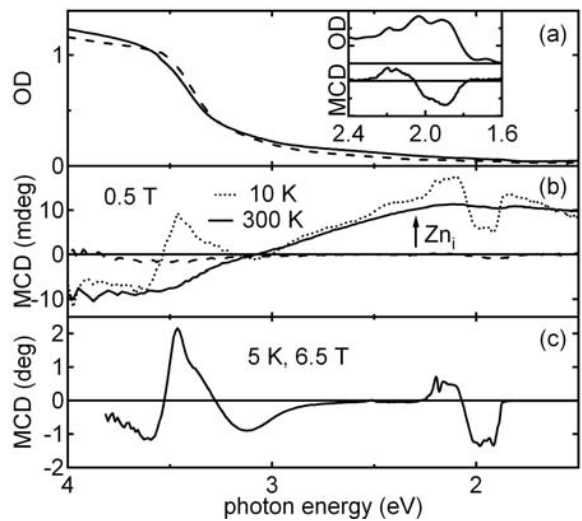

FIG 1. (a) 300 K absorption and (b) 300 and 10 K, 0.5 T MCD spectra of 4.2% $Co^{2+}$:ZnO films. The dashed and solid lines represent before and after exposure to Zn vapor at 450 °C for 1 hr in vacuum, respectively. Inset: Magnified 300 K absorption and MCD spectra of the $^4A_2 \rightarrow {}^4T_1(P)$ ligand field band in paramagnetic $Co^{2+}$:ZnO prior to exposure to Zn vapor. (c) 5 K, 6.5 T MCD of 3.6% $Co^{2+}$:ZnO.

---

* Electronic mail: Gamelin@chem.washington.edu

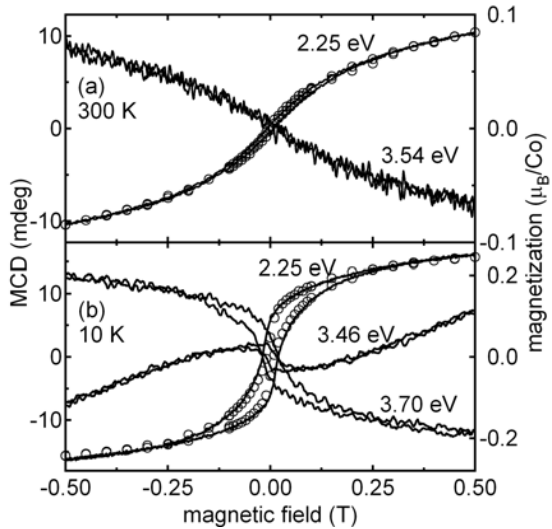

FIG 2. (a) 300 and (b) 10 K magnetic hysteresis loops measured for the 4.2% $Co^{2+}$:ZnO film from Fig. 1 exposed to Zn vapor for 1 hr at 450 °C. SQUID magnetization data (○) are superimposed upon the MCD data (—).

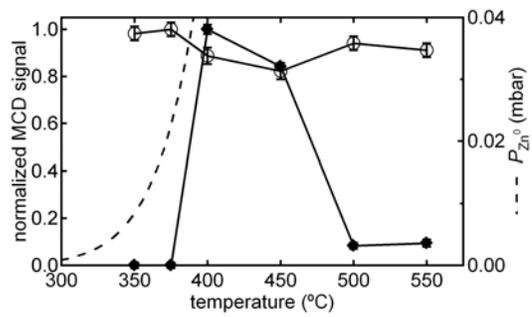

FIG 3. Dependence of the deconvoluted ferromagnetic (■) and paramagnetic (○) MCD intensities of a $Co^{2+}$:ZnO film on Zn vapor anneal temperature, $T_A$. MCD intensities were measured at 300 K and 0.6 T. The dashed line shows the vapor pressure of Zn metal in the same temperature range.

MCD intensities growing in at low temperatures following the anticipated $S = 3/2$ Brillouin magnetization until very low temperatures, consistent with previous measurements.[7,8] Figure 1c shows the 5 K, 6.5 Tesla MCD spectrum of paramagnetic $Co^{2+}$:ZnO. In addition to the ligand-field and excitonic transitions, a valence-band-to-$Co^{2+}$ charge-transfer band is observed at 3.1 eV.[12] Upon exposure of the paramagnetic $Co^{2+}$:ZnO films to Zn vapor, the 300 K MCD intensity was strongly enhanced and its band shape changed (Fig. 1b). The $^4A_2 \rightarrow {}^4T_1(P)$ transition was still weakly evident, superimposed upon a new MCD feature spanning the entire visible range. In contrast with the MCD intensities of the paramagnetic $Co^{2+}$:ZnO, which are strongly temperature dependent, this new MCD signal is nearly temperature independent (Fig. 1b). There were no detectable MCD signals in undoped ZnO films with or without exposure to Zn vapor.

The magnetic field dependence of the 300 and 10 K MCD intensities at different photon energies were measured for several $Co^{2+}$:ZnO films before and after exposure to Zn vapor (Fig. 2). Before exposure to Zn vapor, the 300 K MCD intensities increased linearly with increasing magnetic field (data not shown), consistent with paramagnetism. In contrast, the magnetic field dependence of the MCD spectrum measured after Zn-vapor diffusion revealed magnetic ordering at 300 K. Magnetic hystereses at 300 and 10 K measured by MCD throughout the visible energy range for the film from Fig. 1 are shown in Fig. 2. For comparison, 300 and 10 K magnetization data measured for the same film using a SQUID magnetometer are also included in Fig. 2. The SQUID and MCD hystereses are in excellent agreement. The same hystereses and temperature dependence observed over the entire spectral range suggest that the broad MCD signal in Fig. 1b all comes from a single chromophore, which we propose to be the ferromagnetic DMS $Co^{2+}$:ZnO.

Figure 3 plots the deconvoluted ferromagnetic and paramagnetic $Co^{2+}$:ZnO MCD intensities versus Zn vapor annealing temperature ($T_A$). At $T_A \leq 375$ °C, vacuum annealing with Zn metal had no effect on $Co^{2+}$:ZnO magnetism or MCD. Increasing $T_A$ just slightly from 375 to 400 °C had a dramatic effect on both. The sharp increase in MCD intensity with such a small change in $T_A$ reveals a discrete thermal activation barrier. This barrier is undoubtedly associated with the vaporization of Zn metal. The vapor pressure of Zn metal over this same temperature range[20] is included in Fig. 3 and correlates very well with the activation of ferromagnetism in the $Co^{2+}$:ZnO films. Rather surprisingly, however, the ferromagnetic MCD intensity reached a maximum at $T_A \approx 425$ °C and actually decreased substantially at higher $T_A$. Although the decrease in ferromagnetic MCD intensity with greater $T_A$ at first suggests film decomposition, the MCD spectra of these samples still clearly showed the signature $^4A_2 \rightarrow {}^4T_1(P)$ ligand-field feature of substitutional $Co^{2+}$ in ZnO, indicating that the DMS remained intact under these conditions. Indeed, careful examination of the paramagnetic $Co^{2+}$ ligand-field MCD intensity shows a dependence on $T_A$ that is inversely related to the ferromagnetic MCD signal intensity, having a minimum precisely where ferromagnetism is maximized, and returning again at higher $T_A$ (Fig. 3). These data demonstrate conclusively that cobalt segregation out of the ZnO lattice is not the cause of the diminished ferromagnetism at $T_A \geq 500$ °C and is not the source of the ferromagnetism at $T_A < 500$ °C. These conclusions are supported by several control experiments described previously,[10] in which it was demonstrated, for example, that these films show no detectable cobalt migration even after 15 hrs of vacuum annealing at 500 °C and that films annealed in Zn vapor at 600 °C for 2 hrs are returned quantitatively to their original paramagnetic phase with brief, aerobic annealing (e.g., ~1 min at 500 °C), both inconsistent with phase segregation. The cause of the maximum in ferromagnetism at $T_A \approx 425$ °C remains unknown, but these data clearly support the assignment of the broad MCD signal in Fig. 1b to ferromagnetic $Co^{2+}$:ZnO.

Similar broad sub-bandgap ferromagnetic MCD features have been reported previously for $Co^{2+}$:ZnO[13] and $Co^{2+}$:TiO$_2$.[14,15] Since the ferromagnetic MCD signal of $Co^{2+}$:ZnO did not resemble that of paramagnetic $Co^{2+}$:ZnO there was speculation that it should be attributed to phase-segregated precipitates.[13] In $Co^{2+}$:TiO$_2$, the coincidence of the ferromagnetic MCD hystereses with those measured by the anomalous Hall effect was used to argue an intrinsic origin.[15] Broad sub-bandgap ferromagnetic MCD features have also been observed in Mn:GaAs,[21] Cr:ZnTe,[22] and Mn:InAs,[23] however, materials for which the intrinsic origins of ferromagnetism are less controversial. We propose that the new ferromagnetic MCD signal in Fig. 1b is indeed intrinsic in origin, and that it differs from the MCD of paramagnetic

**Scheme 1:**

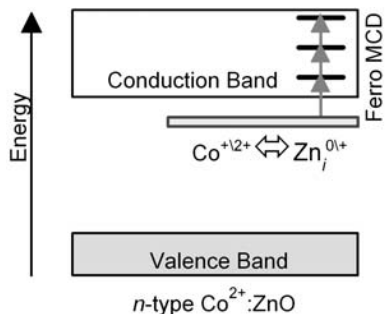

$Co^{2+}$:ZnO because the electronic transitions originate from a fundamentally different chromophore having a very different ground state.

In models of donor-mediated ferromagnetism in $Co^{2+}$:ZnO,[4] shallow donors hybridize with the dopant 3d orbitals to yield a ferromagnetic impurity band near the Fermi level, just below the conduction band.[10] Shallow donor photoionization transitions in ZnO are well known, including the near infrared absorption observed upon doping with $Zn_i$.[10,19] We propose that the broad MCD signal in Fig. 1b arises from analogous low-energy photoionization processes originating from the spin-split donor impurity band, as summarized in Scheme 1. From the breadth of the MCD signal in Fig. 1b it is evident that a multitude of excited states are accessed at similar relatively low photon energies, consistent with this assignment. An alternative assignment that cannot be ruled out involves transitions within an exchange-split spin manifold of the ferromagnetic ground state, in which case the energies of the MCD transitions would relate to the ferromagnetic exchange resonance stabilization energy.[24]

There is extensive literature precedence for the appearance of new spectroscopic transitions upon formation of new magnetic ground states due to exchange coupling. One particularly clear and relevant example is found in the spectroscopy of the well-defined molecular species $[Fe_2(OH)_3(tmtacn)_2]^{2+}$ (tmtacn = trimethyltriazocyclononane), in which two $Fe^{3+}$ ions are coupled together ferromagnetically by an additional itinerant electron via Zener's double exchange mechanism.[25] The spectroscopy of this dimer also bears little resemblance to those of analogous $Fe^{2+}$ or $Fe^{3+}$ complexes, or of analogous antiferromagnetically coupled dimers.[26] Specifically, an intense new absorption and MCD band dominates the visible energy region in spectra of $[Fe_2(OH)_3(tmtacn)_2]^{2+}$ that has no analog in comparable $Fe^{2+}$ or $Fe^{3+}$ complexes. The observation of an MCD spectrum dominated by new transitions in ferromagnetic $Co^{2+}$:ZnO is therefore consistent with formation of a new, easily photoionized ferromagnetic ground state via donor-mediated exchange coupling.

In summary, 300 K ferromagnetism was observed by MCD spectroscopy in $Co^{2+}$:ZnO films exposed to Zn vapor. Optically detected magnetic hystereses showed the same temperature dependence as detected by SQUID magnetization measurements. The sub-bandgap ferromagnetic MCD signal is attributed to low-energy photoionization transitions originating from the new shallow ferromagnetic ground state, transitions that have no analog in paramagnetic $Co^{2+}$:ZnO. A similar interpretation is proposed for the sub-bandgap MCD intensity of ferromagnetic $Co^{2+}$:TiO_2. These MCD data are thus consistent with proposals of donor-mediated ferromagnetism in oxide DMSs.

Financial support from the National Science Foundation (DMR-0239325 and ECS-0224138), the NSF-IGERT program at U.W. (W.K.L.), and the Dreyfus Foundation is gratefully acknowledged. D.R.G. is a Cottrell Scholar of the Research Corporation.

[1] J. K. Furdyna and J. Kossut, in *Semiconductors and Semimetals*, edited by R. K. Willardson and A. C. Beer (Academic, N.Y., 1988), Vol. 25.
[2] S. A. Wolf, D. D. Awschalom, R. A. Buhrman, J. M. Daughton, S. von Molnár, M. L. Roukes, A. Y. Chthelkanova, and D. M. Treger, Science **294**, 1488 (2001); H. Ohno, F. Matsukura, and Y. Ohno, JSAP International **5**, 4 (2002).
[3] For recent reviews see; T. Fukumura, H. Toyosaki, and Y. Yamada, Semicond. Sci. Technol. **20**, S103 (2005); S. J. Pearton, W. H. Heo, M. Ivill, D. P. Norton, and T. Steiner, Semicond. Sci. Technol. **19**, R59 (2004).
[4] J. M. D. Coey, M. Venkatesan, and C. B. Fitzgerald, Nat. Mater. **4**, 173 (2005).
[5] J. H. Park, M. G. Kim, H. M. Jang, S. Ryu, and Y. M. Kim, Appl. Phys. Lett. **84**, 1338 (2004); D. H. Kim, J. S. Yang, K. W. Lee, S. D. Bu, T. W. Noh, S.-J. Oh, Y.-W. Kim, J.-S. Chung, H. Tanaka, H. Y. Lee, and T. Kawai, Appl. Phys. Lett. **81**, 2421 (2002).
[6] K. R. Kittilstved, N. S. Norberg, and D. R. Gamelin, Phys. Rev. Lett. **94**, 147209 (2005); H. Saeki, H. Tabata, and T. Kawai, Sol. State Comm. **120**, 439 (2001).
[7] W. H. Brumage, C. F. Dorman, and C. R. Quade, Phys. Rev. B **63**, 104411 (2001); C. N. R. Rao and F. L. Deepak, J. Mater. Chem. **15**, 573 (2005); S. Kolesnik, B. Dabrowski, and J. Mais, J. Appl. Phys. **95**, 2582 (2004); A. S. Risbud, N. A. Spaldin, Z. Q. Chen, S. Stemmer, and R. Seshadri, Phys. Rev. B **68**, 205202 (2003).
[8] A. C. Tuan, J. D. Bryan, A. B. Pakhomov, V. Shutthanandan, S. Thevuthasan, D. E. McCready, D. Gaspar, M. H. Engelhard, J. W. Rogers, Jr., K. Krishnan, D. R. Gamelin, and S. A. Chambers, Phys. Rev. B **70**, 054424 (2004).
[9] K. Rode, A. Anane, R. Mattana, J.-P. Contour, O. Durand, and R. LeBourgeois, J. Appl. Phys. **93**, 7676 (2003).
[10] D. A. Schwartz and D. R. Gamelin, Adv. Mater. **16**, 2115 (2004).
[11] K. Ando, H. Saito, Z. Jin, T. Fukumura, M. Kawasaki, Y. Matsumoto, and H. Koinuma, Appl. Phys. Lett. **78**, 2700 (2001).
[12] D. A. Schwartz, N. S. Norberg, Q. P. Nguyen, J. M. Parker, and D. R. Gamelin, J. Am. Chem. Soc. **125**, 13205 (2003).
[13] K. Ando, H. Saito, V. Zayets, and M. C. Debnath, J. Phys. Cond. Matter. **16**, S5541 (2004); K. Ando, cond-mat/0208010.
[14] T. Fukumura, Y. Yamada, K. Tamura, K. Nakajima, T. Aoyama, A. Tsukazaki, M. Sumiya, S. Fuke, Y. Segawa, T. Chikyow, T. Hasegawa, H. Koinuma, and M. Kawasaki, Jpn. J. Appl. Phys. **42**, L105 (2003).
[15] H. Toyosaki, T. Fukumura, Y. Yamada, and M. Kawasaki, Appl. Phys. Lett. **86**, 182503 (2005).
[16] Nanostructured films (0.05 - 3.00 μm thick) of $Co^{2+}$:ZnO were prepared on quartz substrates and characterized as described previously.[10] Exposure to Zn vapor was achieved by sealing the films and Zn powder in glass tubes under vacuum (~$10^{-3}$ Torr) and annealing for 1 hr at temperatures between 350 and 550 °C. MCD, absorption, and magnetic susceptibility data were collected as described previously.[12]
[17] H. A. Weakliem, J. Chem. Phys. **36**, 2117 (1962).
[18] D. G. Thomas, Phys. Chem. Solids **3**, 229 (1957).
[19] R. M. de la Cruz, R. Pareja, R. González, L. A. Boatner, and Y. Chen, Phys. Rev. B **45**, 6581 (1992).
[20] *CRC Handbook of Chemistry and Physics* (CRC Press, Boca Raton, FL).
[21] K. J. Yee, R. Chakarvorty, W. L. Lim, X. Liu, M. Kutrowski, L. V. Titova, T. Wojtowicz, J. K. Furdyna, and M. Dobrowolska, J. Supercon. **18**, 131 (2005).
[22] H. Saito, V. Zayets, S. Yamagata, and K. Ando, Phys. Rev. Lett. **90**, 207202 (2003).
[23] K. Ando and H. Munekata, J. Magn. Magn. Mater. **272-276**, 2004 (2004).
[24] R. Yamamoto, Y. Morimoto, and A. Nakamura, Phys. Rev. B **61**, R5062 (2000).
[25] D. R. Gamelin, E. L. Bominaar, M. L. Kirk, K. Wieghardt, and E. I. Solomon, J. Am. Chem. Soc. **118**, 8085 (1996).
[26] A. B. P. Lever, *Inorganic Electronic Spectroscopy* (Elsevier, Amsterdam, 1984, and references therein).